# Hadamard product nonlinear formulation of Galerkin and finite element methods

W. Chen

Permanent mail address: Dr. Wen CHEN, P. O. Box 2-19-201, Jiangshu University of Science & Technology, Zhenjiang City, Jiangsu Province 212013, P. R. China

Present mail address (as a JSPS Postdoctoral Research Fellow): Apt.4, West 1st floor, Himawari-so, 316-2, Wakasato-kitaichi, Nagano-city, Nagano-ken, 380-0926, JAPAN

E-mail: chenw@homer.shinshu-u.ac.jp

Permanent email: chenwwhy@hotmail.com

**Abstract**

A novel nonlinear formulation of finite element and Galerkin methods is presented here, which leads to the Hadamard product expression of the resultant nonlinear algebraic analogue. The presented formulation attains the advantages of weak formulation in the standard finite element and Galerkin schemes and avoids the costly repeated numerical integration of the Jacobian matrix via the recently developed SJT product approach. This also provides possibility of the nonlinear decoupling computations.

integration, Hadamard product, Jacobian matrix, SJT product.

1. **Introduction**

The finite element and Galerkin methods are currently the standard numerical technique in use to solve various nonlinear problems. The methods retain the advantages of weak formulations, which lowers the continuity requirements of matching elements and permits to use simple basis functions. However, these methods demand a great amount of numerical integration computing effort in updating Jacobian matrix of each Newton-Raphson iteration. It was reported that numerical integration often occupied nearly 80% CPU time in the FE and Galerkin solution of large nonlinear systems [1].

Recently, the present author [2, 3] applied the Hadamard product to express the nonlinear formulations of the finite difference (FE) and collocation (pseudo-spectral) methods in an explicit matrix form. Moreover, the SJT product was therein introduced to evaluate the Jacobian matrix efficiently and accurately. A nonlinear decoupling technqiue was also developed by means of the Hadamard and SJT product approach [3, 4]. The simplicity and efficiency of such techniques were well demonstrated through numerical experiments of some benchmark problems. In contrast, when the FE, Galerkin, BE, and least square method are applied to nonlinear problems, the corresponding nonlinear analogue formulations can not be expressed as simple and explicit matrix form of the Hadamard product. It is well known that all these numerical methods have their roots on the weighted residuals method. The objective of this paper is to introduce a novel methodology of Galerkin and finite element methods, which holds the merits of these numerical techniques but overcomes the above-mentioned weaknesses.

## 2. FE and Galerkin nonlinear formulation of Hadamard product

The method of weighted residuals (MWR) can be recognized the origin of almost all popular numerical techniques [5, 6]. Consider the differential equations of the form

$$\psi\{u\} - f = 0, \text{ in } \Omega \tag{1}$$

with the following boundary conditions

$$u = \bar{u}, \quad \text{on} \quad \Gamma_1 \tag{2a}$$

$$q = \partial u/\partial n = \bar{q}, \quad \text{on} \quad \Gamma_2 \tag{2b}$$

where n is the outward normal to the boundary, $\Gamma = \Gamma_1 + \Gamma_2$, and the upper bars indicate known values. More complex boundary conditions can be easily included but they will not be considered here for the sake of brevity. In the MWR, the desired function u in the differential governing equation is first approximated by a set of linearly independent basis functions $\phi_k(x)$, such that

$$u = \hat{u} = \sum_{j=1}^{n} c_j \phi_j, \tag{3}$$

where $c_k$'s are the unknown parameters. In the Galerkin and FE methods, the basis functions are usually chosen so as to satisfy certain given conditions such as the boundary conditions and the degree of continuity. In addition, these basis functions should be complete.

Substituting equation (3) into equation (1) produces an error, which is called the residual, namely,

$$\psi\{\hat{u}\} - f = R \neq 0. \tag{4}$$

This error or residual R is forced to be zero in the average sense by setting weighted integral of the residuals equal to zero, namely,

$$\int_{\Omega}[\psi\{\hat{u}\} - f]W_j d\Omega = \int_{\Omega} RW_j d\Omega = 0, \quad j=1,2,\ldots,N, \tag{5}$$

where $W_j$'s are weighting functions. The differences among weighting functions and basis functions give rise to different numerical techniques such the Galerkin, least square, finite element, boundary element, spectral methods, finite difference and collocation methods.

This paper places its emphasis on the nonlinear computations. Let us consider the quadratic nonlinear operator of the form:

$$p(u)q(u) + L(u) = f, \tag{6}$$

where p(u), q(u) and L(u) are linear differential operators, f is the constant. The traditional scheme of weighted residuals approximates equation (6) by

$$\int_{\Omega}[p(\hat{u})q(\hat{u}) + L(\hat{u}) - f]W_j d\Omega = 0, \quad j=1,2,\ldots,N, \tag{7}$$

where $W_j$ denotes weight function in MWR. Here we present an innovative scheme

$$\int_\Omega p(\hat{u})W_j d\Omega \int_\Omega q(\hat{u})W_j d\Omega + \int_\Omega [L(\hat{u}) - f]W_j d\Omega = 0, \quad j=1,2,\ldots,N \qquad (8)$$

different from the standard equation (7). It is noted that the key distinction of equations (7) and (8) lies in that latter weights linear operators p() and q() separately. The nonlinear operators can be understood certain combinations of linear operators. Before further development, we first introduce the concepts of Hadamard matrix product, power and function.

**Definition 2.1** Let matrices A=[$a_{ij}$] and B=[$b_{ij}$]$\in C^{N \times M}$, the Hadamard product of matrices is defined as A°B= [$a_{ij} b_{ij}$]$\in C^{N \times M}$. where $C^{N \times M}$ denotes the set of N×M real matrices.

The weighting residuals of operators p(), q() and L() in equations (8) can be expressed as

$$\int_\Omega p\{\hat{u}\}W_j d\Omega = Ax, \qquad (9a)$$

$$\int_\Omega q\{\hat{u}\}W_j d\Omega = Bx, \qquad (9b)$$

$$\int_\Omega L\{\hat{u}\}W_j d\Omega = Dx, \qquad (9c)$$

where $x = \{c_i\}$, $c_i$'s are the undetermined parameters in equation (3). Therefore, it is intuitively found that the nonlinear formulations of weighted residuals equation (8) can be expressed as

$$(Ax) \circ (Bx) + Dx = b, \qquad (10)$$

where b is the constant vector. For the traditional scheme of weighted residual (equation

(7)), the nonlinear formulation can be only expressed in matrix form as [2, 3]

$$D_{n \times n} x + G_{n \times n^2}(x \times x) = b \qquad (11)$$

by using theorem (2.1), where D is given in equation (9c),

$$G_{n \times n^2} = \int_{\Omega} [p(\hat{u}) \otimes q(\hat{u})] W_j d\Omega \in C^{n \times n^2}. \qquad (12)$$

The preceding inferences implicitly assume that the basis functions in equation (3) satisfy all the boundary conditions. However, this is not necessary in general. Among all methods originated from the method of weighted residuals, the more interesting for engineering applications is the FE and Galerkin methods. The residuals of these two methods are weighted by the basis functions of the approximate solution, and the boundary conditions can be found by integrating the governing equations by parts, which leads to the so-called weak formulations. By weighting the residuals of governing equation (1) and boundary conditions (2a, b) via basis functions, we have

$$\int_{\Omega} [p(\hat{u})q(\hat{u}) + L(\hat{u}) - f] \phi_j d\Omega = \int_{\Gamma_2} (q - \bar{q}) \phi_j d\Gamma - \int_{\Gamma_1} (u - \bar{u}) \frac{\partial \phi_j}{\partial n} d\Gamma \qquad (13)$$

in the standard way. In contrast, we have

$$\int_{\Omega} p(\hat{u}) \phi_j d\Omega \int_{\Omega} q(\hat{u}) \phi_j d\Omega + \int_{\Omega} [L(\hat{u}) - f] \phi_j d\Omega = \int_{\Gamma_2} (q - \bar{q}) \phi_j d\Gamma - \int_{\Gamma_1} (u - \bar{u}) \frac{\partial \phi_j}{\partial n} d\Gamma \qquad (14)$$

in the present way. Formulation (14) can be expressed in the Hadamard product form as in equation (10).

The given innovative scheme has the important features of requiring the formulation of all linear differential operators only once, which save considerable computing resources by avoiding the repeated integration of Jacobian matrix in the iterative solution of nonlinear systems. In addition, due to the Hadamard form formulation available in the present WRM nonlinear discretization, simple iteration method (Picard method) can be effectively used to solve these nonlinear algebraic equations as pointed out in Chen [3]. The present strategy is also much simpler to use. More importantly, possible benefits includes the rapid evaluation of Jacobian matrix via SJT product and nonlinear decoupling computations [2-4].

### 3. Remarks

One of the major factors which affects the efficiency of the FE and Calerkin methods is need to repeat numerical integration of Jacobian (stiffness) matrices. The presented formulation cures this deficiency and yet they maintain strong geometric and boundary flexibility with the weak formulation. The conventional nonlinear FE method may also be too complex mathematically for routine applications. In contrast, the present Hadamard product formulation is an explicit and simple matrix analysis. More importantly, the SJT product and the relative decoupling algorithms are now capable of being extended to the FE and Galerkin solution of nonlinear problems. References [2-4] provided some benchmark numerical examples to demonstrate the superiority of these approaches.